\documentclass[apjl]{emulateapj}
\usepackage{natbib}
\usepackage{hyperref}
\hypersetup{
    colorlinks=true,
    linkcolor=blue,
    citecolor=blue,
    filecolor=blue,
    urlcolor=blue
}
 \providecommand{\adsurl}[1]{\href{#1}{ADS}}
\usepackage{amsmath,amssymb}
\usepackage{mathptmx}
\usepackage{mathtools}
\DeclareMathAlphabet{\pazocal}{OMS}{zplm}{m}{n}

\usepackage{graphicx}

\begin{document}

\title{Could Solar Radiation Pressure Explain `Oumuamua's Peculiar Acceleration?}
 \author{Shmuel Bialy$^\star$ and Abraham Loeb}
 \affiliation{Harvard Smithsonian Center for Astrophysics, 60 Garden st., Cambridge, MA, 02138
 }
 \email{$^\star$SBialy@cfa.harvard.edu}
\slugcomment{Accepted for publication in the Astrophysical Journal Letters}

\begin{abstract}
 `Oumuamua (1I/2017 U1) is the first object of interstellar origin observed in the Solar System.
 Recently, \citet{Micheli2018} reported that `Oumuamua showed deviations from a Keplerian orbit at a high statistical significance. 
 The observed trajectory is best explained by an excess radial acceleration $\Delta a \propto r^{-2}$, where $r$ is the distance of `Oumuamua from the Sun.
Such an acceleration is naturally expected for comets, driven by the evaporating material.
However, recent observational and theoretical studies imply that `Oumuamua is not an active comet.
We explore the possibility that the excess acceleration results from Solar radiation pressure.
The required mass-to-area ratio is $(m/A)\approx 0.1$ g cm$^{-2}$. For a  thin sheet this requires a thickness of $\approx 0.3-0.9$ mm.
We find that although extremely thin, such an object  would  survive an interstellar travel over Galactic distances of $\sim 5$ kpc, withstanding collisions with gas and dust-grains as well as stresses from rotation and tidal forces.
We discuss the possible origins of such an object.
Our general results apply to any light probes designed for interstellar travel.

\end{abstract}

\section{Introduction}
\label{sec: intro}
On October 19, 2017, the first interstellar object in the Solar System, `Oumuamua (1I/2017 U1)  was discovered by the PAN-STARRS1 survey.
It has a highly hyperbolic trajectory (with eccentricity $e=1.1956 \pm 0.0006$) and pre-entry velocity of $v_{\infty}\approx 26$ km s$^{-1}$ \citep{Meech2017}.
Based on the survey properties and the single detection, \citet{do2018} estimated the interstellar density of objects like `Oumuamua or larger to be $n \approx 2 \times 10^{15}$ pc$^{-3}$, 2-8
orders of magnitude larger than expected by previous theoretical models \citep{Moro-Martin2009}.
The large variations in its apparent magnitude and the non-trivial periodicity of the lightcurve, suggest that `Oumuamua is rotating in an excited spin state (tumbling motion), and has an extreme aspect ratio of at least $5:1$   \citep{Fraser2018, Drahus2018}, an unprecedented value for previously known asteroids and comets in the Solar System.
\citet{Belton2018} have shown that if `Oumuamua rotates in its highest rotational energy state, it should be extremely oblate (pancake-like).

Recently, \citet{Micheli2018} reported the detection of non-gravitational acceleration in the motion of `Oumuamua, at a statistical significance of 30$\sigma$. Their best-fit to the data is obtained for a model with a non-constant excess acceleration which scales with distance from the Sun, $r$, as $\Delta a \propto r^{-2}$, but other power-law index values are also possible.
They concluded that the observed acceleration is most likely the result of a cometary activity.
Yet, despite its close Solar approach of $r = 0.25$ AU, `Oumuamua shows no signs of a any cometary activity, no cometary tail, nor gas emission/absorption lines were observed \citep{Meech2017, Knight2017, Jewitt2017, Ye2017, Fitzsimmons2017}.
From  a theoretical point of view, \citet{Rafikov2018} has shown that if outgassing was responsible for the acceleration \citep[as originally proposed by][]{Micheli2018}, then the associated outgassing torques would have driven a rapid evolution in `Oumuamua's spin, incompatible with observations.

If not cometary activity, what can drive the non-gravitational acceleration observed?
In this {\it Letter} we explore the possibility of `Oumuamua being a thin object accelerated by Solar radiation pressure, which would naturally result in an excess acceleration $\Delta a\propto r^{-2}$. 
\footnote{Interestingly, a similar approach was adopted historically regarding the anomalous orbit of Phobos \citep{Shklovskii1962}.}
However, for radiation pressure to be effective, the mass-to-area ratio must be very small.
In \S \ref{sec: rad pressure} we derive the required mass-to-area ratio and find $(m/A)\approx 0.1$ g cm$^{-2}$, corresponding to an effective thin sheet of thickness $w \approx 0.3-0.9$ mm.
We explore the ability of such an unusually thin object to survive interstellar travel, considering collisions with interstellar dust and gas (\S \ref{sec: travel}), as well as to withstand the tensile stresses caused by rotation and tidal forces (\S \ref{sec: strength}).
Finally, in \S \ref{sec: discussion} we discuss the possible implications of the unusual requirements on the shape of `Oumuamua.

\section{Acceleration by Radiation Pressure}
\label{sec: rad pressure}
\citet{Micheli2018} had shown that `Oumuamua
experiences an excess radial acceleration, with
their best fit model
\begin{align}
\label{eq: a_obs}
    \Delta a &= a_0 \left( \frac{r}{\rm AU} \right)^{n} \\ \nonumber
    {\rm with} \ \ n &= -2 \ , \\
     a_0 & = (4.92 \pm 0.16) \times 10^{-4} \ {\rm cm \ s^{-2}} \ .
\end{align}
The value for $a_0$ is averaged over timescales much longer than `Oumuamua's rotation period.

An acceleration of this form is naturally produced by radiation pressure,
\begin{equation}
\label{eq: rad_press}
P = C_R \frac{L_{\odot}}{4 \pi r^2 c} \ ,
\end{equation}
where $L_{\odot}$ is the Solar luminosity, $c$ is the speed of  light, and $C_R$ is a coefficient of order unity which depends on the object's composition and geometry.
For a sheet perpendicular to the Sun-object vector $C_R= 1+\epsilon$ where $\epsilon$ is the reflectivity. For a perfect reflector $\epsilon=1$, and $C_R=2$, whereas for a perfect absorber $\epsilon = 0$ and $C_R = 1$.

For an object of mass $m$ and area $A$
the acceleration would be
\begin{align}
\label{eq: a_rad_press}
a &=  \frac{PA}{m} 
= \left( \frac{L_{\odot}}{4 \pi r^2 c} \right)
 \left(\frac{A}{m}  \right)
  C_R \\ \nonumber
& = 4.6 \times 10^{-5} 
\left( \frac{r}{\rm AU}     \right)^{-2} 
\left( \frac{m/A}{\rm g \ cm^{-2} } \right)^{-1}
C_R \
{\rm cm \ s^{-2} }
\ .
\end{align}
Comparing Eq.~(\ref{eq: a_obs}) and (\ref{eq: a_rad_press}) we find that the requirement on the mass-to-area ratio is
\begin{equation}
\label{eq: constraint  m_A}
     \left( \frac{m}{A} \right) 
    = (9.3 \pm 0.3) \times 10^{-2} \ C_R \ {\rm g \ cm^{-2}} \ .
\end{equation}
For a planar body with mass density $\rho$, this translates into a requirement on the body's thickness
\begin{equation}
w \ = \ \frac{m}{A \rho} = (9.3 \pm 0.3) \times 10^{-2} \rho_0^{-1} \ C_R \ {\rm cm} \ , 
\end{equation}
where $\rho_0=\rho/(10^0 \ {\rm g \ cm^{-3}})$. Typically, $\rho_0 \approx 1-3$, giving a thin sheet of 0.3--0.9 mm thickness.
Other geometries are also possible, and are discussed in \S \ref{sec: discussion}. 
The force exerted by the Solar wind on a solid surface is negligible compared to that of the Solar radiation field and is neglected hereafter.


The observed magnitude of `Oumuamua constrains its area to be
$A \approx 8 \times 10^6 \alpha^{-1}$ cm$^2$, where $\alpha$ is the albedo \citep{Jewitt2017}.
This corresponds to an effective radius, $R_{\rm eff} \equiv \sqrt{A}/\pi = 16 \alpha^{-1/2}$ meters.
For our estimation of the mass-to-area ratio, this area translates into a mass of $m \approx 740  (C_R/\alpha)$ kg.

\section{Maximum  Distance for Interstellar Travel}
\label{sec: travel}
\begin{figure}[t]
	\centering
	\includegraphics[width=0.5\textwidth]{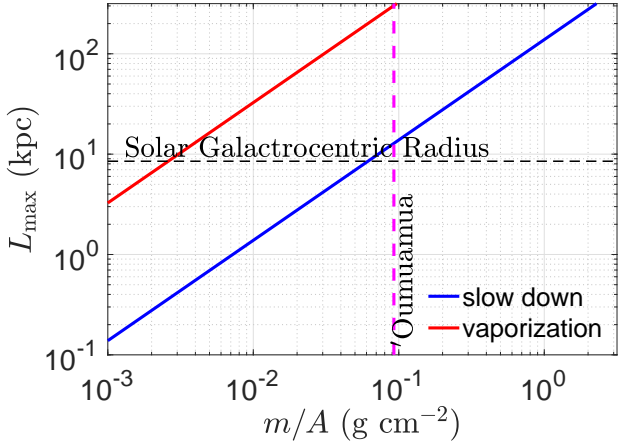} 
	\caption{
	 The maximum allowed travel distance through the interstellar medium (ISM), as a function of $(m/A)$. The blue and red lines are limitations obtained by slow-down due to gas accumulation, and vaporization by dust-collisions, respectively.
The plotted results are for a mean ISM proton density of $\langle n \rangle \sim 1$ cm$^{-3}$. All lines scale as $1/\langle n \rangle$.
The dashed magenta line is our constraint on `Oumuamua based on its excess acceleration.
The Solar Galactrocentric distance is also indicated.
		}
		\label{fig: L_max}
\end{figure}

Next we explore the implications of impacts with interstellar dust-grains and gas particles, in terms of momentum and energy transfer. 
We obtain general requirements for the object's mass-to-area ratio, or alternatively, for the maximum interstellar distance that can be traveled before it encounters appreciable slow-down or evaporation.

\subsection{Momentum Transfer - Slow Down}
An object with a cross sectional area $A$ traveling a distance $L$ through the ISM would accumulate an ISM gas mass of,
\begin{equation}
\label{eq: M ISM gas swept}
M_{\rm gas} = A \Sigma_{\rm gas} = 1.4 m_{p} \langle n \rangle L A \ ,
\end{equation}
where $\Sigma_{\rm gas}$ is the accumulated mass column density of interstellar gas,  $m_p$ is the proton mass, $\langle n \rangle$ is the mean proton number density averaged along the object's trajectory, and the factor 1.4 accounts for the contribution of helium to the mass density of the ISM.
For trajectories that span Galactic distances the contribution of the Solar System to the accumulated column is negligible.
The contribution of accumulated dust to the momentum transfer is also negligible, because the typical dust-to-gas mass ratio in the Galaxy is $\approx 1/100$.

Once $M_{\rm ISM}$ approaches the object's mass, $m$, the momentum of the traveling object will decrease by a significant amount.
The requirement $M_{\rm ISM}/m \ll 1$ translates into a maximum allowed value on the object's mass-to-area ratio, giving
\begin{align}
\label{eq: m_A momentum}
\left( \frac{m}{A} \right)_{\rm min, p} &= \Sigma_{\rm gas} = 1.4 m_p \langle n \rangle L \\ \nonumber
& = 7.2 \times 10^{-3} \langle n \rangle_0 L_{0} \ {\rm g \ cm^{-2}} \ .
\end{align}
In the last equality we normalized to the typical values $\langle n \rangle_0 = \langle n \rangle/(10^0 \ {\rm cm^{-3}})$ and  $L_0 = L/(10^0 \ {\rm kpc})$. 


Given a mass-to-area ratio, the maximum travel distance is 
\begin{align}
\label{eq: L_max p}
L_{\rm max, p} &=  \frac{1}{1.4 m_p \langle n \rangle} \left( \frac{m}{A} \right)\\ \nonumber
& = 14 \langle n \rangle_0^{-1} 
\left( \frac{m}{A} \right)_{-1}
 \ {\rm kpc} \ .
\end{align} 
In the second equation we denoted 
$(m/A)_{-1} \equiv (m/A)/(10^{-1} {\rm \ g \ cm^{-2}})$.
Figure~\ref{fig: L_max} shows the results from Eq.~(\ref{eq: L_max p}) as a function of ($m/A$).
The dashed vertical line indicates our constraint on ($m/A$) for `Oumuamua (Eq.~\ref{eq: constraint  m_A}).
Evidently, `Oumuamua can travel Galactic distances before encountering appreciable slow-down.



\subsection{Energy Transfer - Collisions with Dust-Grains}

Collisions with dust grains at high velocities will induce crater formation by melting and evaporation of the target material.
Since the typical time between dust collisions is long compared to the solidification time, any molten material will solidify before the next collision occurs, and thus will only cause a deformation of the object's surface material, not reduction in mass.
On the other hand, atoms vaporized through collisions can escape and thus cause a mass ablation.

We would like to estimate the minimum mass-to-area ratio required for the object to not lose significant fraction of its mass upon dust-grain collisions.
Let $m_d$ be the colliding dust-grain mass, and $\phi$ the fraction of the kinetic energy that is converted into vaporization of the object's body. The total number of vaporized atoms per collision is then,
\begin{equation}
\label{eq: dust melting one grain}
    N_v =  \frac{m_v}{\bar{m}} = \phi \frac{ m_d v^2}{2U_v}\ ,
\end{equation}
where $m_v$ is the mass of vaporized material in the object, $\bar{m}$ is the mean atomic mass of the object, and $U_v$ is the vaporization energy.
Although highly simplistic, this analysis captures the results of the detailed  theoretical model of \citet{Tielens1994} and the  empirical data from   \citet{Okeefe1977}.
A good match to the numerical results is obtained for $\phi = 0.2$.

Over a distance $L$ there will be many collisions.
Adopting the conservative assumption that for each collision all the vaporized material escapes to the ISM, we can account for all the collisions along a path-length, $L$, by replacing $m_d$ in Eq.~(\ref{eq: dust melting one grain}) with the total accumulated dust mass, 
\begin{equation}
\label{eq: total dust}
    M_{d} = \Sigma_{\rm dust} A = \Sigma_{\rm gas} \varphi_{dg} A = 1.4 m_p \langle n \rangle L \varphi_{dg} A \ ,
\end{equation}
where $\Sigma_{\rm dust}$ is the dust column density and $\phi_{dg}$ is the dust-to-gas mass ratio. This gives,
\begin{equation}
\label{eq: dust melting all dust}
    \frac{m_v}{\bar{m}} = \phi \frac{ 1.4 m_p \langle n \rangle L \varphi_{dg} A v^2}{2U_v} \ .
\end{equation}
Requiring that the total vaporized mass not exceed half of the object's mass, we obtain a constraint on the minimum mass-to-area ratio of the object, 
\begin{align}
\label{eq: m_A dust melt}
\left( \frac{m}{A} \right)_{\rm min, m} 
&= 1.4 m_p \langle n \rangle L 
\left( \frac{\phi  \varphi_{dg} \bar{m} v^2}{U_v} \right)  \\ \nonumber
& = 3.1 \times 10^{-4} \langle n \rangle_0 L_{0}  
\left( \frac{\varphi_{-2}\bar{m}_{12} v^2_{26}}{U_{4}} \right) \ {\rm g \ cm^{-2}} \ .
\end{align}
Here we defined the normalized parameters, $\varphi_{-2}=\varphi_{dg}/10^{-2}$, $\bar{m}_{12}=\bar{m}/(12 m_p)$, as appropriate for carbon based materials (e.g., graphite or diamond); $U_{4}=U/(4 \ {\rm eV})$, as appropriate for typical vaporization energies (e.g., for graphite, $U_v=4.2$ eV); and $v_{26}=v/(26 \ {\rm km \ s^{-1}})$, the velocity at infinity of `Oumuamua.
For a given mass-to-area ratio, the maximum allowed distance before significant evaporation is
\begin{align}
\label{eq: L_max melt}
    L_{\rm max, d} &= \frac{1}{1.4 m_p \langle n \rangle} \left( \frac{U_m}{\phi  \varphi_{dg} \bar{m} v^2} \right) \left( \frac{m}{A} \right) \\ \nonumber
    & = 330 \langle n \rangle_0^{-1} \left( \frac{U_{4}}{\varphi_{-2}\bar{m}_{12} v^2_{26}} \right) \left( 
\frac{m}{A} 
\right)_{-1} \ {\rm kpc} \ .
\end{align}
For our constrained value for the mass-to-area ratio, `Oumuamua can travel through the entire galaxy before a significant fraction of its mass is evaporated.
Evaporation becomes important  at higher speeds.
Comparing Eqs.~(\ref{eq: m_A momentum}) and (\ref{eq: m_A dust melt}) we find that only for speeds above
\begin{align}
    v_{\rm crit} &= \sqrt{\frac{2U_m}{\bar{m}\varphi \phi}} \\ \nonumber
    & = 130 \left( \frac{U_{4}}{\varphi_{-2}\bar{m}_{12} } \right)^{1/2} \ {\rm km \ s^{-1}} \ ,
\end{align}
vaporization dominates over slow-down.


\subsection{Energy Transfer - Collisions with Gas Particles}

When an object travels at a high speed, collisions with atoms in the ISM can potentially transfer sufficient energy to produce sputtering. 
This process was studied in the context of dust grains  in hot shocks \citep{Tielens1994}.
For an object traveling at a velocity $v$, over a distance $L$ through the ISM, the total number of sputtered particles is,
\begin{equation}
    N_s  = \frac{m_s}{\bar{m}} =  2 A  L \langle n \rangle Y_{\rm tot} \ ,
\end{equation}
where $Y_{\rm tot}=\sum_i Y_i x_i$ is the total sputtering yield (which depends on the kinetic energy), summed over collisions with different species (i.e., H, He, and metals), and $x_i$ is the abundance of the colliding species  relative to hydrogen.

The minimal mass-to-area ratio below which half of the object's mass will be sputtered is,
\begin{align}
\label{eq: m_A_sputtering}
\left( \frac{m}{A} \right)_{\rm min, s} &= \bar{m} \langle n \rangle L  Y_{\rm tot}  \\ \nonumber
&= 6.2 \times 10^{-5} \bar{m}_{12}  \langle n \rangle_0 L_0 Y_{-3} \ {\rm g \ cm^{-2}} \ .
\end{align}
In the second equality we normalized to $Y_{\rm tot}=10^{-3}$, corresponding to kinetic energies $E \approx 30 - 100$ eV,  corresponding to $v \approx 40 - 70$ km s$^{-1}$.
For lower speeds, as that of `Oumuamua, the yield is even lower, further decreasing the value of $(m/A)_{\rm min, s}$.
At higher speeds, the yield increases but typically remains below $0.01$ \citep{Tielens1994}, thus at any velocity, vaporization and slow-down remain the dominating processes limiting the allowed distance an object can travel through the ISM.

Cosmic-rays are expected to cause even less damage.
Although their energy density is comparable to that of the ISM gas, they deposit only a very small fraction of their energy as they penetrate through the thin object.

\section{Tensile Stresses}
\label{sec: strength}

A thin object can be torn apart by 
centrifugal forces or tidal forces
if its tensile strength is not sufficiently strong.
Typical values for the tensile strengths of various materials are shown in Table \ref{table: tensile strength}.
Next, we calculate whether centrifugal or tidal forces can destroy `Oumuamua.

\begin{table}[]
\caption{Tensile Strengths}
\centering 
\begin{tabular}{l l}
\hline\hline 
material  & tensile strength (${\rm dyne \  cm^{-2}}$) 
  \\ [0.5ex]
  \hline 
  67P/Churyumov-Gerasimenko$^1$ & 10-50 
\\
Meteorites$^2$ & $(1 - 5) \times 10^7$ 
\\
Iron & $3 \times 10^{7}$ \\
  Diamond & $2 \times 10^{10}$ \\
  Silicon (monocrystalline) & $7 \times 10^{10}$ \\
  \hline
    \hline
$^1$ \citet{Attree2018} & \\
$^2$ \citet{Petrovic2001} & \\
\end{tabular}
\label{table: tensile strength} 
\end{table}

\subsection{Rotation}
\label{sub: rot}

Oumuamua's lightcurve shows periodic modulations on an order of 6-8 hours. Ignoring the tumbling motion, let us estimate the tensile stress originating from the centrifugal force.
The largest stress is produced if the object is elongated such that the longest dimension is perpendicular to the rotation axis. We denote this dimension as $d$.
Considering the object as made of two halves, each located with a center of mass at a distance $d/4$ from the rotation axis, and ignoring self gravity, a radial force of magnitude,
\begin{equation}
F = \frac{1}{2} m \Omega^2 \frac{d}{4} \ ,
\end{equation}
will be exerted on each half.
The associated tensile stress is,
\begin{align}
P_{\rm rot} &= \frac{1}{4} \rho d^2  \Omega^2\\ \nonumber \nonumber
&= 0.25 \ \rho_0 \ d_{4}^2 \ \Omega_{-4}^2 \  {\rm dyne \ cm^{-2}} \ ,
\end{align}
where $d_4 \equiv d/(10^4 \ {\rm cm})$, $\Omega_{-4} \equiv \Omega/(10^{-4} \ {\rm s^{-1}})$.
This is much smaller than typical tensile strengths of normal materials, and even of that of the comet 67P/Churyumov-Gerasimenko (see Table \ref{table: tensile strength}).
Thus, even when self-gravity is ignored, `Oumuamua can easily withstand its centrifugal force.

\subsection{Tidal Forces}
The tidal force will be maximal if the long dimension of the object is parallel to the Sun-object vector.
Again, modeling the object as consisting of two halves as in \S \ref{sub: rot},
the difference in the gravitational force experienced by the far and near ends of the object is,
\begin{equation}
\delta F \approx \frac{1}{4} d \frac{GM_{\odot}m}{r^3} \ ,
\end{equation}
where $r$ is the distance of the center of mass from the Sun.
The associated tensile stress,
\begin{align}
P_{\rm tid} &\approx \frac{1}{4} \rho d^2 \frac{GM_{\odot}}{r^3} \\ \nonumber
& = 9.9 \times 10^{-7} \ \rho_0 \ d_4^2 \left(\frac{r}{{\rm AU}} \right)^{-3} \ {\rm dyne \ cm^{-2}} \ .
\end{align}
Even at perihelion ($r=0.25$ AU), the tensile stress is negligible.

The critical distance below which tidal forces dominate over centrifugal is
\begin{align}
R_{\rm tid} = \left(\frac{G M}{\Omega^2} \right)^{1/3}  = 3.4  \Omega_{-4}^{-2/3} \left(  \frac{M}{M_{\odot}}\right)^{1/3} \ R_{\odot} \ .
\end{align}
Thus, unless `Oumuamua encountered an extremely close approach to a star in its past, it is unlikely that tidal forces played any significant role.

\section{Summary and Discussion}
\label{sec: discussion}
We have shown that the observed non-gravitational acceleration of `Oumuamua, may be explained by Solar radiation pressure.
This requires a small mass-to-area ratio for `Oumuamua of $(m/A) \approx 0.1$ g cm$^{-2}$.
For a planar geometry and typical mass densities of 1--3 g cm$^{-2}$  this gives an effective thickness of only 0.9--0.3 mm, respectively. For a material with lower mass density, the inferred effective thickness is proportionally larger. 
We find that although very thin, such an object can travel over galactic distances, maintaining its momentum and withstanding collisional destruction by dust-grains and gas,
as well as centrifugal and tidal forces.
For `Oumuamua, the limiting factor is the slow-down by accumulated ISM mass, which limits its maximal travel distance to $\sim 10$  kpc (for a mean ISM particle density of $\sim 1$ cm$^{-3}$).
Our inferred thin geometry is consistent with studies of its tumbling motion. In particular, \citet{Belton2018} inferred that `Oumuamua is likely to be an extremely oblate spheroid (pancake) assuming that it is  excited by external torques to its highest energy state.

While our scenario may naturally explains the peculiar acceleration of `Oumuamua,
it opens up the question what kind of object might have such a small mass-to-area ratio?
The observations are not sufficiently sensitive to provide a resolved image of `Oumuamua, and one can only speculate on its possible geometry and nature. 
Although periodic variations in the apparent magnitude are observed, there are still too many degrees of freedom (e.g., observing angle, non-uniform reflectively, etc.) to definitely constrain the geometry.
The geometry should not necessarily be that of a planar sheet, but may acquire other shapes, e.g., involving a curved sheet, a hollow cone or ellipsoidal, etc.
Depending on the geometry our estimated value for the mass-to-area ratio will change (through $C_R$ in Eq.~\ref{eq: constraint  m_A}), but the correction is typically of order unity.

Known Solar System objects, like asteroids and comets have mass-to-area ratios orders of magnitude larger than our estimate for `Oumuamua.
If radiation pressure is the accelerating force, then `Oumuamua represents a new class of thin interstellar material, either  produced naturally,through a yet unknown process in
the ISM or in proto-planetary disks, or of an artificial origin.

Considering an artificial origin, one possibility is that `Oumuamua is a lightsail,   floating in interstellar space
as a debris from an advanced technological equipment  \citep{Loeb2018}.
Lightsails with similar dimensions have been designed and constructed by our own civilization, including the IKAROS project and the Starshot Initiative\footnote{A list of books and papers on lightsails is provided in \href{http://breakthroughinitiatives.org/research/3}{http://breakthroughinitiatives.org/research/3}. The IKAROS project is discussed in  \href{http://global.jaxa.jp/projects/sat/ikaros/}{http://global.jaxa.jp/projects/sat/ikaros/}.}. 
The lightsail technology might be abundantly used for transportation of cargo between planets \citep{Guillochon2015} or between stars \citep{Lingam2017}. In the former case, dynamical ejection from a planetary System could result in space debris of equipment that is not operational any more
\footnote{Note that `Oumuamua was found not to show any radio emission down to a fraction of the power of a cell phone transmission \citep{Enriquez2018a, Tingay2018, Harp2018}.}
\citep{Loeb2018}, and is floating at the characteristic speed of stars relative to each other in the Solar neighborhood.
This would account for the various anomalies of `Oumuamua, such as the 
unusual geometry
 inferred from its lightcurve \citep{Meech2017, Fraser2018, Drahus2018, Belton2018}, 
its low thermal emission, suggesting high reflectivity \citep{Trilling2018},
and its deviation from a Keplerian orbit \citep{Micheli2018} without any sign of a cometary tail \citep{Meech2017, Knight2017, Jewitt2017, Ye2017, Fitzsimmons2017} or spin-up torques \citep{Rafikov2018}.
 Although `Oumuamua has a red surface color, similar to organic-rich surfaces of Solar-System comets and D-type asteroids \citep{Meech2017}, this does not contradict the artificial scenario, since irrespective of the object's composition, as it travels through the ISM its surface will be covered by a layer of interstellar dust, which is itself composed of organic-rich materials \citep{Draine2003}.

Alternatively, a more exotic scenario is that 
`Oumuamua may be a fully operational probe sent {\it intentionally} to Earth vicinity by an alien civilization.
Based on the PAN-STARRS1 survey characteristics, and assuming natural origins following {\it random} trajectories, 
  \citet[][]{do2018} derived that the interstellar number density of `Oumuamua-like objects should be extremely high, $\sim 2 \times 10^{15}$ pc$^{-3}$, equivalent to $\sim 10^{15}$  ejected planetisimals  per star, and a factor of 100 to 10$^8$ larger than predicted by
  theoretical models \citep{Moro-Martin2009}.
This discrepancy is readily solved if `Oumuamua does not follow a random trajectory but is rather a targeted probe.
Interestingly, `Oumuamua's entry velocity is found to be extremely close to the velocity of the Local Standard of Rest, in a kinematic region that is occupied by less than 1 to 500 stars \citep{Mamajek2017}.

Since it is too late to image `Oumuamua with existing telescopes or chase it with chemical propulsion rockets \citep[][but see \citealt{Hein2017}]{Seligman2018}, its likely origin and mechanical properties could only be deciphered by searching for other objects of its type in the future. In addition to the vast unbound population, 
thousands of interstellar `Oumuamua-like space-debris
 are expected to be trapped at any given time in the Solar System through gravitational interaction with Jupiter and the Sun \citep{Lingam2018}. 
Deep wide-area surveys of the type expected with the Large Synoptic Survey Telescope (LSST)\footnote{See  https://www.lsst.org/} will be particularly powerful in searching for additional members of `Oumuamua's population of objects.

A survey for lightsails as technosignatures in the Solar System is warranted, irrespective of whether `Oumuamua is one of them.

\acknowledgements
We thank Manasvi Lingam, Paul Duffell, Quanzhi Ye, and an anonymous referee for helpful comments.
This work was supported in part by a grant from the Breakthrough Prize Foundation.

\end{document}